\documentclass[%
 reprint,
 amsmath,amssymb,
 aps,
showkeys
]{revtex4-1}
\usepackage{graphicx}
\usepackage{dcolumn}
\usepackage{bm}
\usepackage{verbatim}
\usepackage{soul,xcolor}
\usepackage{amsmath,amstext}
\usepackage{amsmath,amsthm,amsfonts,amssymb}
\usepackage{mathtools,amssymb}

\let\Im\undefined

\DeclareMathOperator{\Im}{\mathrm{Im}}

\usepackage{hyperref}

\begin{document}

\preprint{APS/123-QED}

\title{Black String Thermal Equilibrium}

\author{João Chakrian}%
 \email{joao.chakrian@ufrpe.br}

\author{Antônio de Pádua Santos}
\email{antonio.padua@ufrpe.br}

\affiliation{%
 Departamento de Física, Universidade Federal Rural de Pernambuco, 52171-900 Recife, PE, Brazil
 }%

\date{\today}

\begin{abstract}
The black string solution is a model of black holes in (3+1) dimensions that can be related to the BTZ (2+1) dimensions black hole solution. This object can be seen as a cylindrical black hole. In this work it is considered the black string Hawking temperature obtained by the Hamilton-Jacobi method and a description of a system composed by a black string and a thermal reservoir, considering the thermal equilibrium state.
\end{abstract}


\maketitle

\section{\label{sec:intro} Introduction}

In General Relativity, the spherically symmetric black hole solutions are widely studied due to their relevant role in describing various physical systems of astrophysical and cosmological importance, such as in the description of the final state of a gravitational collapse of stars. The interest of describing black holes with different topology is not considered due the formulation of hoop conjecture, established by Thorne \cite{Thorne}, which states that event horizons can be created only if a mass is compressed into a region with circumference less than $4\pi GM$ in every direction, which means that only spherically symmetric black holes should exist. However, the hoop conjecture holds only when cosmological constant is assumed to be zero, and therefore, if one admits negative cosmological constant, the spacetime shall become asymptotically Anti-de Sitter. This fact makes possible to investigate cylindrically symmetric black hole solutions, that is, black strings. 

The black strings thermodynamic properties are known \cite{Chakrian,Saifullah-thermo} and in particular the heat capacity positive behavior is interesting for this paper once heat capacity, and as consequence the specific heat, is always positive. This fact can be understood as the property of a system composed by the black string and a thermal reservoir reach a thermal equilibrium state. In this work, it is analyzed the black string behavior when thermal equilibrium is considered. It is assumed that the black string mass per unit of length changes due Hawking radiation since its temperature (Hawking temperature) is different from the temperature of the thermal reservoir $T_R$. In this conjecture, the black string temperature can be higher, $T > T_R$, or lower $T < T_R$ than the reservoir temperature. The temporal evolution of mass per unit of length is investigated in this work.

\section{\label{sec:black_string} Black string}

The papers \cite{Lemos-bs,Lemos-rotating-bs} performed by Lemos were a breakthrough in the subject of cylindrical black holes since he has shown that the black strings are predicted when a negative cosmological constant is considered, and moreover, it is possible \cite{Lemos-btz-bs} to relate this (3+1) dimensions solution to the BTZ (2+1) dimension solution. In such initial papers, the black string solution arises from solving the motion equations that result from the Einstein-Hilbert four dimensions action

\begin{equation} \label{eq:E-H_action}
S = \frac{1}{16\pi G}\int \sqrt{-\mathrm{g}}(R - 2\Lambda)d^4x,
\end{equation}
where $S$ is the Einstein-Hilbert action in four dimensions, $\textsl{g}$ is the determinant of the metric tensor, $\textsl{g}=\text{det}(g_{\mu\nu})$, $R$ is the Ricci scalar, $\Lambda$ is the cosmological constant and $G$ is the Newtonian gravitational constant. The black string solution, considering $G = \hbar = c = 1$, is given by \cite{Lemos-rotating-bs}

\begin{eqnarray} \label{eq:BS_metric}
ds^2 &=& -\left(\alpha^2r^2 - \frac{4M}{\alpha r}\right)dt^2 + \left(\alpha^2r^2 - \frac{4M}{\alpha r}\right)^{-1}dr^2 \nonumber\\
&+& r^2d\phi^2 + \alpha^2r^2dz^2,
\end{eqnarray}
where $-\infty < t < \infty$, the radial coordinate $0\leq r < \infty$, the angular coordinate $0 \leq \phi < 2\pi$, the axial coordinate $-\infty < z < \infty$, $\alpha^2 = -\frac{\Lambda}{3} > 0$ and $M$ the mass per unit length in the $z$-direction. This solution represents a static straight black string with one event horizon at 

\begin{equation} \label{eq:horizon}
    r_+ = \frac{(4M_+)^{\frac{1}{3}}}{\alpha}.
\end{equation}

\section{\label{sec:temperature} Temperature}

The Hamilton-Jacobi method represents an approach to investigate the quantum tunneling from scalar field in black string spacetime. This method is based on the tunneling approach by Parikh and Wilczeck \cite{Parikh}, the path integrals method by Srinivasan \cite{Srinivasan-complex} and can be found in the references \cite{Saifullah-emission, Saifullah-quantum}. We start from the Klein-Gordon equation on curved spaces for a scalar field $\Phi$

\begin{equation} \label{eq:K-G}
    \mathrm{g}^{\mu\nu}\partial_\mu\partial_\nu\Phi - \frac{m^2}{\hbar^2}\Phi = 0,
\end{equation}
where $\mathrm{g}^{\mu\nu}$ is the metric tensor related to the line element \eqref{eq:BS_metric} and $m$ is the mass of the field. In order to solve this equation, we use the ansatz

\begin{equation} \label{eq:WKB}
    \Phi(t,r,\phi,z) = \exp\left\{\frac{i}{\hbar}S(t,r,\phi,z)\right\},
\end{equation}
where $S$ represents the trajectory action associated to the particle tunneling. We can recognize this solution as the WKB approximation \cite{Birrell}. This low wavelength solution is justified since the amount of null geodesics tends to infinity at the event horizon which means that there is a blueshift \cite{Parikh}. Substituting \eqref{eq:WKB} into \eqref{eq:K-G} and using the first term of the expansion of the action in terms of $\hbar$, we find

\begin{equation}
    \mathrm{g}^{tt}(\partial_t S)^2 + \mathrm{g}^{rr}(\partial_r S)^2 + \mathrm{g}^{\phi\phi}(\partial_\phi S)^2 + \mathrm{g}^{zz}(\partial_z S)^2 + m^2 = 0.
    \label{eq:action_part}
\end{equation}
The solution to equation \eqref{eq:action_part} can be considered by separation of variables as follows

\begin{equation}
    S(t,r,\phi,z) = - Et + W(r) + J_\phi \phi + J_z z + C,
\end{equation}
where $E$ is a constant associated to the energy, $W(r)$ is a function to be determined, $J_\phi,J_z$ are constants associated to angular momentum in $\phi$ and $z$ direction, respectively, and $C$ is a constant. After performing an integration on complex plane, the function $W(r)$ is given by

\begin{equation}
    W_{\pm}(r) = \pm \frac{i\pi E}{3\alpha^2 r_+}
\end{equation}
where it was used the relation \eqref{eq:horizon} and we remark that $W_+ = - W_-$ . The probabilities of crossing the event horizon can be written as 

\begin{eqnarray}
    \Gamma_{\mbox{\scriptsize emission}} &=& \exp\left\{- \frac{2}{\hbar}\left[\Im(W_+) + \Im(C)\right]\right\} \label{eq:emissao1} \\
    \Gamma_{\mbox{\scriptsize absorption}} &=& \exp\left\{- \frac{2}{\hbar}\left[\Im(W_-) + \Im(C)\right]\right\}
\end{eqnarray}
and by the condition that the probability of entering in the event horizon of the black string is $100\%$ \cite{Saifullah-emission}, we determine that $\Im(W_-) = \Im(C)$ and the replacement of this result into \eqref{eq:emissao1}  leads to

\begin{eqnarray} \label{eq:emissao}
    \Gamma_{\mbox{\scriptsize emission}} &=& \exp\left\{- \frac{4}{\hbar}\Im(W_+)\right\} \nonumber\\
    \Gamma_{\mbox{\scriptsize emission}} &=& \exp\left\{- \frac{4}{\hbar}\frac{\pi E}{3\alpha^2 r_+}\right\}.
\end{eqnarray}
Now, if one compares the emission factor \eqref{eq:emissao} to the canonical ensemble Boltzmann factor $\exp\left\{E/T_H\right\}$, the Hawking temperature is written as, (we set $\hbar = 1$)

\begin{equation} \label{eq:T_hawking}
    T_H = \frac{3\alpha^2}{4\pi}r_+.
\end{equation}
This result can also be found in the references \cite{Saifullah-thermo, Cai} by a different approach and in the references \cite{Saifullah-emission, Saifullah-quantum} by using the Hamilton-Jacobi method. It is interesting to see that the temperature behavior is linear in terms of the event horizon radius $r_+$. If one uses the relation \eqref{eq:horizon} to write this temperature in terms of mass per unit of length associated to the event horizon radius, one obtains 

\begin{equation} \label{eq:T-M_+}
    T_H = \frac{3\alpha}{2\pi}\left(\frac{M_+}{2}\right)^{\frac{1}{3}},
\end{equation}
which makes it clear that $T_H \rightarrow 0$ only if $M_+ \rightarrow 0$. From this point, unless stated otherwise we shall consider $T_H = T$.

\section{\label{sec:vazio} Empty Space as Thermal Reservoir}

 The Stefan-Boltzmann equation can be used to investigate the energy flux due the difference of temperature of the black string and the empty space as presented in the reference \cite{Santi}. We adapt this procedure in order to study the temporal evolution of the black string mass per unit length. The Stefan-Boltzmann relation can be written as

\begin{equation} \label{eq:def_Stefan-Boltzmann}
    J = \lambda A(T^4 - T_R^4)
\end{equation}
where $T$ is identified, in this work, as the Hawking temperature associated to the black string, $T_R$ is the temperature associated to the reservoir, $L$ represents the energy flux through the system components, $\lambda$ is the Stefan-Boltzmann constant: $\lambda = \pi^2k_B^4/60\hbar^3 c^2$ in SI units and $A$ is the area per unit of length of the black string $A = 2\alpha\pi r_+^2$ \cite{Cai}. The system proposed is defined as the black string, in vacuum, with temperature $T_H$, we use $T_R = 0$ and, for this case, the relation \eqref{eq:def_Stefan-Boltzmann} reduces to 

\begin{equation} \label{eq:S-B_vacuo}
    J = \lambda AT^4.
\end{equation}
It is important to point out that it shall be used arbitrary units because, in the point of view of this work, the general behavior of the results are interesting and their values could be investigated  considering any choice of units. Then, rewriting the relation \eqref{eq:S-B_vacuo} using the equation \eqref{eq:horizon} to write $r_+$ in terms of $M_+$ and considering $k_B = G = c = \hbar = 1$, we have

\begin{equation} \label{eq:L_vacuo}
    J = \frac{27\alpha^3}{160\pi}M_+^2.
\end{equation}
Now, for the investigated system we can consider $J$ as the temporal variation of energy of the black string and, as consequence, its temporal variation of the mass per unit length since the energy is modified by emission (or absorption) of radiation. In this perspective, it is possible to write $J = - dM_+/dt$ and using the relation \eqref{eq:L_vacuo} one finds

\begin{equation} \label{eq:eq-dif-M_vazio}
    \frac{dM_+}{dt} = - \frac{27\alpha^3}{160\pi}M_+^2.
\end{equation}
Integrating this relation, considering the initial conditions $t_0 = 0$ and $M_+(t_0) = M_0$, as

\begin{equation}
    \int_{M_0}^{M} \frac{dM'_+}{{M'_+}^2} = - \frac{27\alpha^3}{160\pi}\int_{t_0=0}^t dt'
\end{equation}
we obtain the temporal evolution of the black string mass per unit length considering the empty space as the thermal reservoir:

\begin{equation} \label{eq:M_vacuo} 
    M(t) = \frac{1}{\displaystyle at + \frac{1}{M_0}}.
\end{equation}
Here, $a = \frac{27\alpha^3}{160\pi}$. This relation establishes that the ``complete evaporation'' ($M \rightarrow 0$) would demand an infinite time. This fact is physically plausible because if $M \rightarrow 0$, then $T \rightarrow 0$ from relation \eqref{eq:T-M_+} and $T \rightarrow 0$ should not happen due the third law of thermodynamics. In other words, it is not possible to the black string completely evaporate. We remark that this prediction suggests a different behavior in comparison to the Schwarzchild black hole since for the last one it is predicted the existence of a lifetime which depends on the initial mass of the black hole.

\begin{figure}[ht]
    \centering
    \includegraphics[scale=.5]{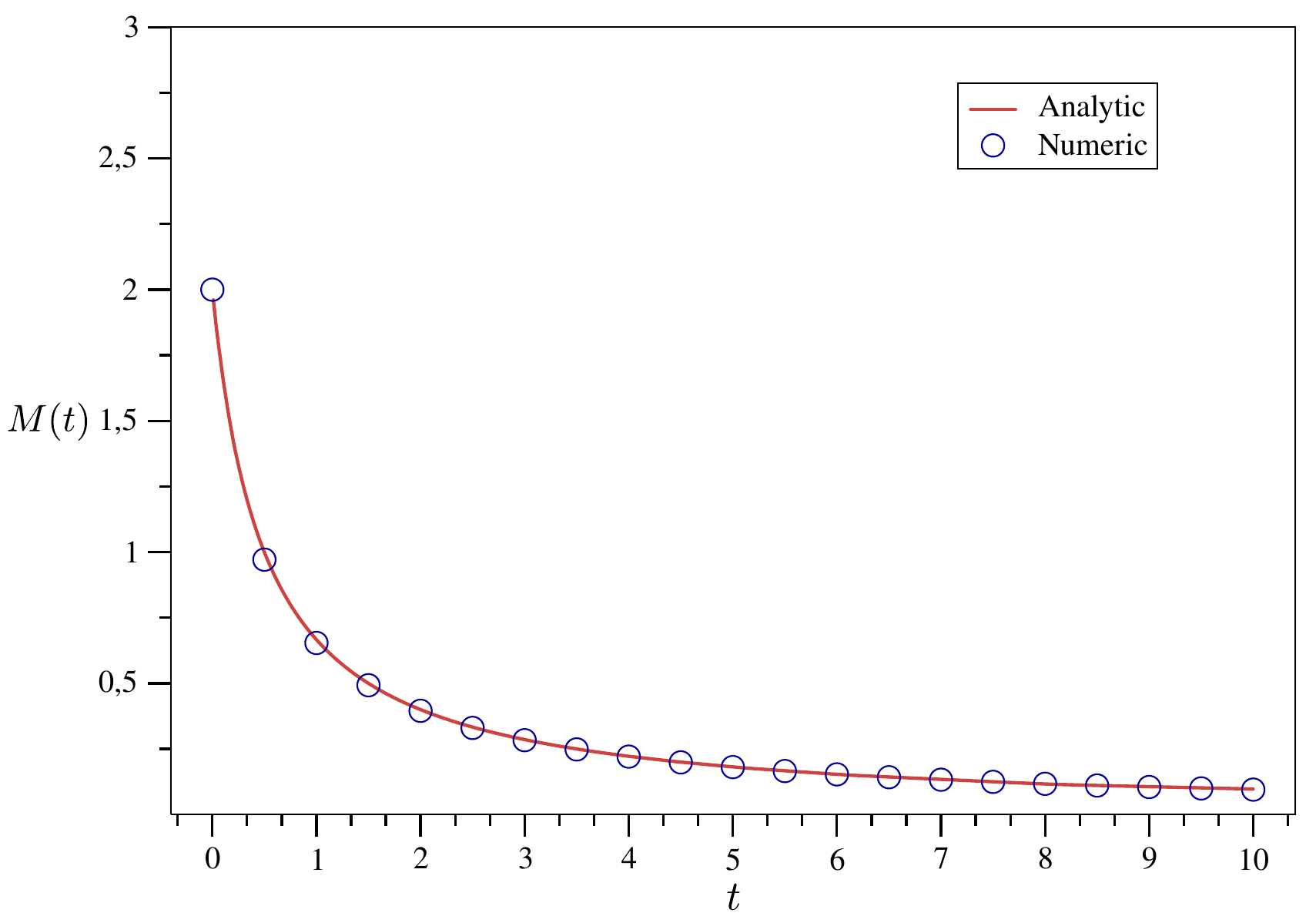}
    \caption{Temporal evolution of the mass per unit of length considering the empty space as thermal reservoir. Here, the parameters are $a = 1$ and $M_0 = 2$ expressed in arbitrary units.}
    \label{fig:Mxt_vazio}
\end{figure}

The figure \ref{fig:Mxt_vazio} presents a plot of the relation \eqref{eq:M_vacuo} which represents the analytic solution to the equation \eqref{eq:eq-dif-M_vazio} and also presents a numerical\footnote{It was applied the Classical Runge-Kutta method using the open source software of mathematics \textit{SageMath}. All numerical calculations presented in this work were performed using the mentioned software.} solution to such equation. It is possible to observe the decrease in the value of the mass per unit length. This fact was expected since the black string temperature is higher than the empty space temperature $T > T_R$ and therefore the black string radiates and evaporates continuously until the system reach the equilibrium temperature.

\section{\label{sec:CMB} Cosmic Microwave Background as Thermal Reservoir}

After the temperature and density decrease due Universe expansion, the interactions among photons and hydrogen atoms also decreased. The photons, earlier with sufficient energy to remove electrons from atoms, became less energetic and could no longer remove electrons, but moved freely without interacting with matter. Hence, the Big Bang model predicts the cosmic radiation background and its energy is associated to the hydrogen atom ionization energy. Such radiation was detected as black body radiation in microwave frequency by Arno Penzias and Robert Wilson in 1965 \cite{Penzias-Wilson, Dicke}. The current approximated value associated to the Cosmic Microwave Background - CMB is $T_{\mbox{\scriptsize CMB}} \approx 2.7255 \mbox{K}$. We consider the CMB as the thermal reservoir and we study the energy exchange in the system. Therefore, it is possible to associate a thermal current (energy flux) by considering the Stefan-Boltzmann relation, writing $T_R = T_{\mbox{\scriptsize CMB}}$, as

\begin{equation}
    J = \lambda A\left(T^4 - T_{\mbox{\scriptsize CMB}}^4\right).
\end{equation}
Performing the procedure adopted in the previous section, the temporal evolution of the black string mass per unit of length can be determined by the following differential equation 

\begin{equation} \label{eq:eq-dif-M_cmb}
    \frac{dM_+}{dt} = - aM_+^2 + bM_+^{\frac{2}{3}},
\end{equation}
where

\begin{equation} \label{eq:def_a_b}
    \begin{aligned}
    a &= \frac{27\alpha^3 }{160\pi} \\
    b &= \frac{4^{\frac{2}{3}}\pi^3}{30\alpha}T_{\mbox{\scriptsize CMB}}^4.
    \end{aligned}
\end{equation}

The relation \eqref{eq:eq-dif-M_cmb}, which represents the temporal evolution of the mass per unit of length considering the CMB as thermal reservoir, can be investigated for the purpose of obtaining a clue with respect to the mass behavior of the system. It is possible to calculate an equilibrium mass relation: after some modifications, we integrate the equation \eqref{eq:eq-dif-M_cmb} as

\begin{equation}
    \int dt' = \int \frac{dM'_+}{- a{M'_+}^2 + b{M'_+}^{\frac{2}{3}}} = - 3 \int \frac{dx}{ax^4 - b}
\end{equation}
where $x = {M'_+}^{\frac{1}{3}}$, $a$ and $b$ are given in equation \eqref{eq:def_a_b}. Performing the integration, we obtain

\begin{eqnarray}
    t &=& \frac{3}{4a^{\frac{1}{4}}b^{\frac{3}{4}}}\left\{2\arctan\left[\left(\frac{a}{b}\right)^{\frac{1}{4}}M^{\frac{1}{3}}\right] + \ln\left(b^{\frac{1}{4}} + a^{\frac{1}{4}}M^{\frac{1}{3}}\right) \right. \nonumber\\
    &-& \left. \ln\left(b^{\frac{1}{4}} - a^{\frac{1}{4}}M^{\frac{1}{3}}\right)\right\} + K
\end{eqnarray}
where $K$ is an integration constant. We notice a change in behavior when the logarithms arguments are zero, that is,

\begin{eqnarray}
    b^{\frac{1}{4}} - a^{\frac{1}{4}}M^{\frac{1}{3}} &=& 0 \\
    b^{\frac{1}{4}} + a^{\frac{1}{4}}M^{\frac{1}{3}} &=& 0.
\end{eqnarray}
The second relation above results negative mass and for this reason such relation is not important. However, the first relation provides the mass associated to the change in system behavior as

\begin{equation} \label{eq:M_equilibrio}
    M_e = \left(\frac{b}{a}\right)^{\frac{3}{4}} = \frac{16}{27}\left(\frac{\pi T_{\mbox{\scriptsize CMB}} }{\alpha}\right)^3.
\end{equation}
The mass per unit of length shown in equation \eqref{eq:M_equilibrio} represents the mass value for which the thermal equilibrium of the system is reached. This interpretation can be confirmed if we calculate what should be the mass associated with the CMB temperature. Considering the relation \eqref{eq:T_hawking} applied for CMB one obtains

\begin{equation*}
    r_{\mbox{\scriptsize CMB}}= \frac{4\pi}{3\alpha^2}T_{\mbox{\scriptsize CMB}}
\end{equation*}
and now, considering the relation \eqref{eq:horizon}, one obtains the following result which confirms the interpretation concerning the equation \eqref{eq:M_equilibrio}

\begin{equation*}
    M_{\mbox{\scriptsize CMB}} = \frac{16}{27}\left(\frac{\pi T_{\mbox{\scriptsize CMB}} }{\alpha}\right)^3 = M_e.
\end{equation*}

\begin{figure}[ht]
    \centering
    \includegraphics[scale=.5]{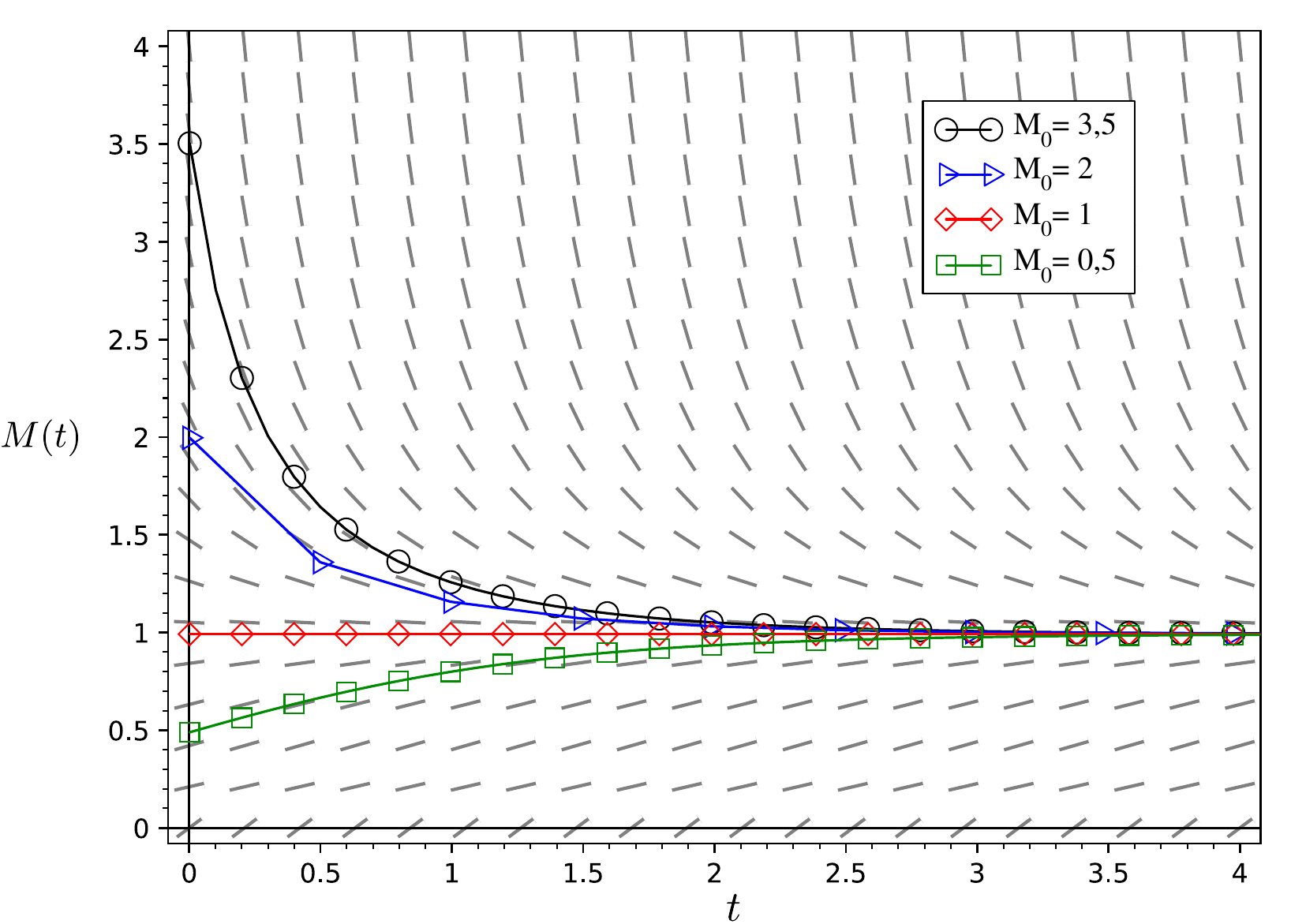}
    \caption{For the solution presented it is considered $a = b = 1$ and some values for the initial mass per unit of length $M_0$ in arbitrary units.}
    \label{fig:Mxt-field_cmb}
\end{figure}

The figure \ref{fig:Mxt-field_cmb} shows a plot of the solution to the equation \eqref{eq:eq-dif-M_cmb}. It can be seen that the predictions provided by the previous discussion in the section \ref{sec:intro} have been confirmed. One can see that the mass per unit of length assumes a constant value, $M = 1$ and such fact corroborates two predictions: first, the mass per unit of length reaches a constant value and it indicates that the temperature gradient ceases and there is no energy exchange between black string and CMB anymore; second, the equilibrium mass per unit of length $M_e = 1$ confirms relation \eqref{eq:M_equilibrio} since, for this plot, it was set $a = b = 1$ and the relation \eqref{eq:M_equilibrio} predicts that, for such choice of parameters, $M_e = 1$.

\section{\label{sec:CMB_evoluc} Cosmic Microwave Background in Evolution as Thermal Reservoir}

The CMB evolves according to the temporal evolution of the universe. We consider this time evolution divided into three eras: the radiation era, the matter era, and the dark energy era. Each one represents the era when there is a predominance of each one of the cosmic fluids (radiation, matter and dark energy) considering the big bang as the beginning of time. The energy exchange between the black string and the CMB is investigated by considering that the temperature $T_{\mbox{\scriptsize CMB}}$ assumes different behaviors for each era of the Universe. The temperature for each era can be written, in SI units, as\footnote{These relations, presented in SI units are known and their demonstration can be found in \cite{Santi}. In this work, we apply such relations in arbitrary units to understand the black string temporal evolution.}

\begin{eqnarray}
    T_R &=& \frac{T_{\mbox{\scriptsize CMB}}}{\displaystyle \left(\frac{32\pi G}{3}\rho_{R_0}\right)^{\frac{1}{4}}t^{\frac{1}{2}}} \label{eq:T_radiation}\\
    T_M &=& \frac{T_{\mbox{\scriptsize CMB}}}{\left(6\pi G \rho_{M_0}\right)^{\frac{1}{3}}t^{\frac{2}{3}}} \label{eq:T_matter}\\
    T_\Lambda &=& T_{\mbox{\scriptsize CMB}}\exp\left[- \left(\frac{8\pi G}{3}\rho_{\Lambda_0}\right)^{\frac{1}{2}}t\right] \label{eq:T_dark}
\end{eqnarray}
where $\rho_{R_0}$,$\rho_{M_0}$, $\rho_{\Lambda_0}$ represent the current densities of radiation, matter and dark energy, respectively. Photons, neutrinos, cosmic microwave background and gravitons are associated to the radiation era; hadronic and non-hadronic matter are associated to the matter era; and the dark energy era is associated to dark energy.

\subsection{\label{subsec:era_rad} Radiation Era} 

\begin{figure}[ht]
    \centering
    \includegraphics[scale=.5]{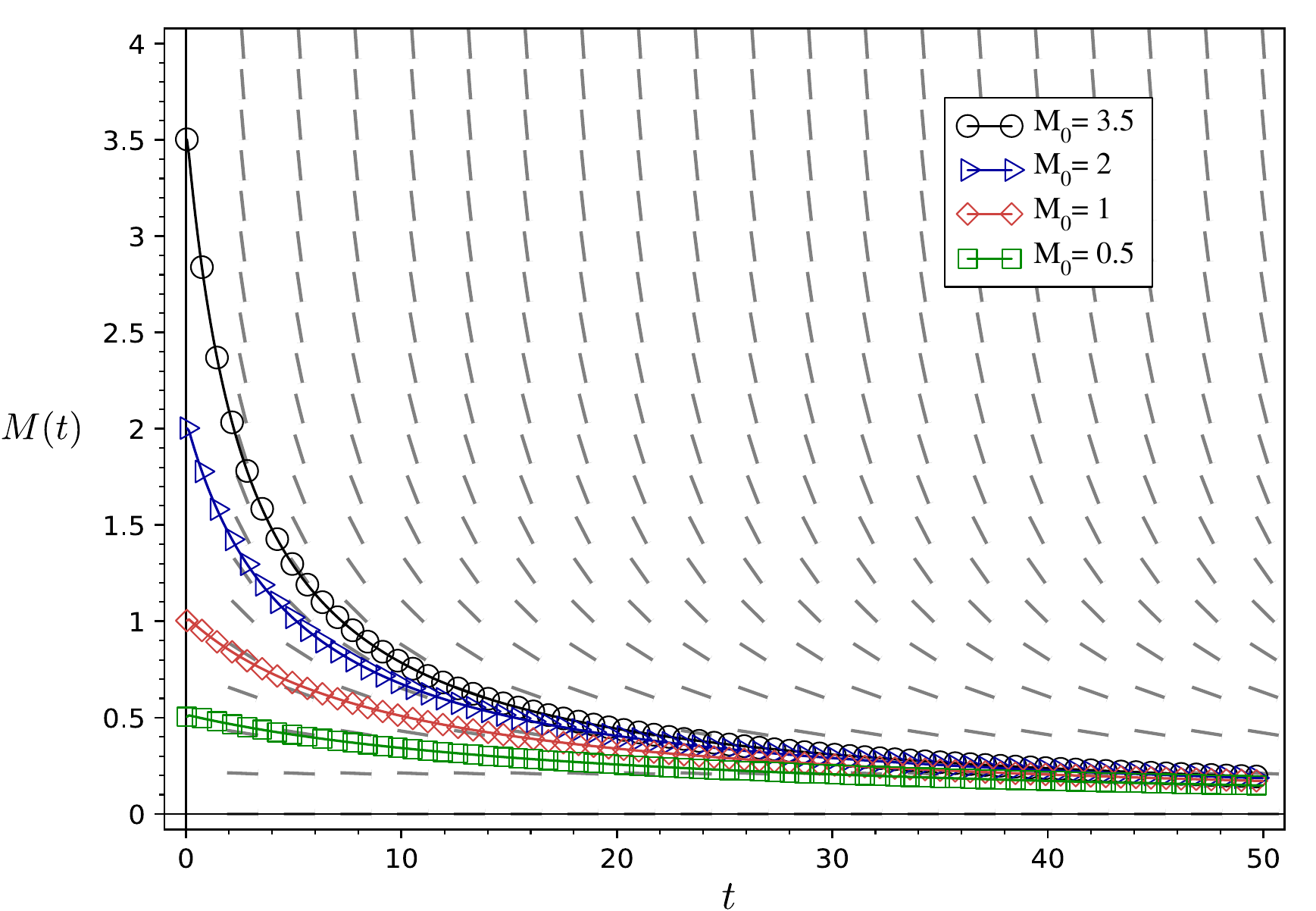}
    \caption{Radiation era. For this plot, $a = 0.1$, $b = \delta_R = 0.001$}
    \label{fig:Mxt-field_rad}
\end{figure}

We analyze the energy flux between the black string and the thermal reservoir by rewriting the Stefan-Boltzmann law for each era as

\begin{equation} \label{eq:Stefan-Boltzmann_evoluc}
    J = \lambda A\left(T^4 - T_i^4\right),
\end{equation}
where $A = 2\alpha\pi r_+^2$, $\lambda$ is the Stefan-Boltzmann constant as introduced previously and $i = R, M, \Lambda$ represents radiation, matter or dark energy, respectively. Applying the relations \eqref{eq:T_radiation} and \eqref{eq:Stefan-Boltzmann_evoluc} we obtain the following differential equation in radiation era

\begin{equation} \label{eq:eq-dif-M_rad}
    \frac{dM_+}{dt} = - aM_+^2 + \frac{b\delta_R}{t^2}M_+^{\frac{2}{3}}
\end{equation}
where 

\begin{equation} \label{eq:delta_R}
    \delta_R = \frac{3}{\displaystyle 32\pi G\rho_{R_0}},
\end{equation}
$a$ and $b$ are given in relation \eqref{eq:def_a_b}. The solution to the equation \eqref{eq:eq-dif-M_rad} is shown in figure \ref{fig:Mxt-field_rad} and it is possible to see curves approach a horizontal asymptote defined by the value of $M_e$, the value of mass per unit length associated to the thermal equilibrium which represents an equilibrium stable point. The decrease in value of mass per unit length represents the black string evaporation by Hawking radiation.

\subsection{\label{subsec:era_mat} Matter Era} 

\begin{figure}[ht]
    \centering
    \includegraphics[scale=.55]{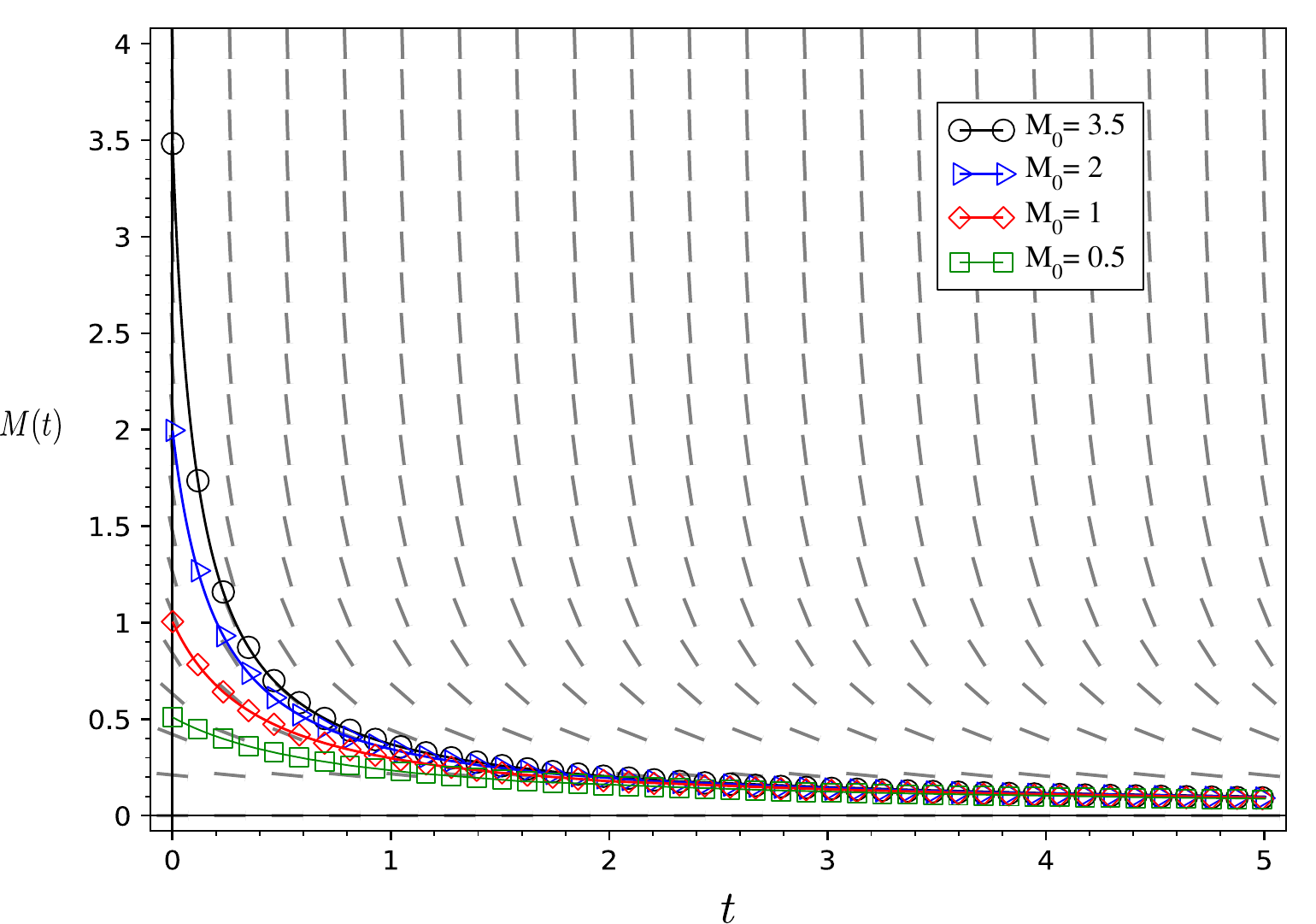}
    \caption{Matter era. For this plot it is considered $a = 2.5$, $b = 10^{-5}$ and $\delta_M = 2\cdot 10^{-4}$.}
    \label{fig:Mxt-field_mat}
\end{figure}

Now we consider the relation \eqref{eq:Stefan-Boltzmann_evoluc} for the matter era. The temporal evolution of mass per unit length can be written as 

\begin{equation} \label{eq:eq-dif-M_matter}
    \frac{dM_+}{dt} = -aM_+^2 + \frac{b\delta_M}{t^{\frac{8}{3}}}M_+^{\frac{2}{3}}
\end{equation}
where

\begin{equation} \label{eq:delta_M}
    \delta_M = \frac{1}{\displaystyle \left(6\pi G\rho_{M_0}\right)^{\frac{4}{3}}}.
\end{equation}
It is possible to observe that, as in the previous cases, the equation \eqref{eq:eq-dif-M_matter} is a non-linear differential equation and the figure \ref{fig:Mxt-field_mat} describes its numerical solution. This graph indicates a decrease its mass value in a similar way to the previous ones. This represents that in the matter era the black string evaporates and decreases its mass per unit length until the thermal equilibrium is established. It is important to remark that the decrease of mass value occurs faster than the other cases. Investigating the horizontal axis, one can observe, by comparison between \ref{fig:Mxt-field_mat} and the previous plots, that the decrease of mass per unit length occurs significantly for values ten times smaller than the previous cases. This fact indicates that there is an intense black string evaporation predicted in matter era.

\subsection{\label{subsec:era_dark} Dark Energy Era} 

\begin{figure}[ht]
    \centering
    \includegraphics[scale=.5]{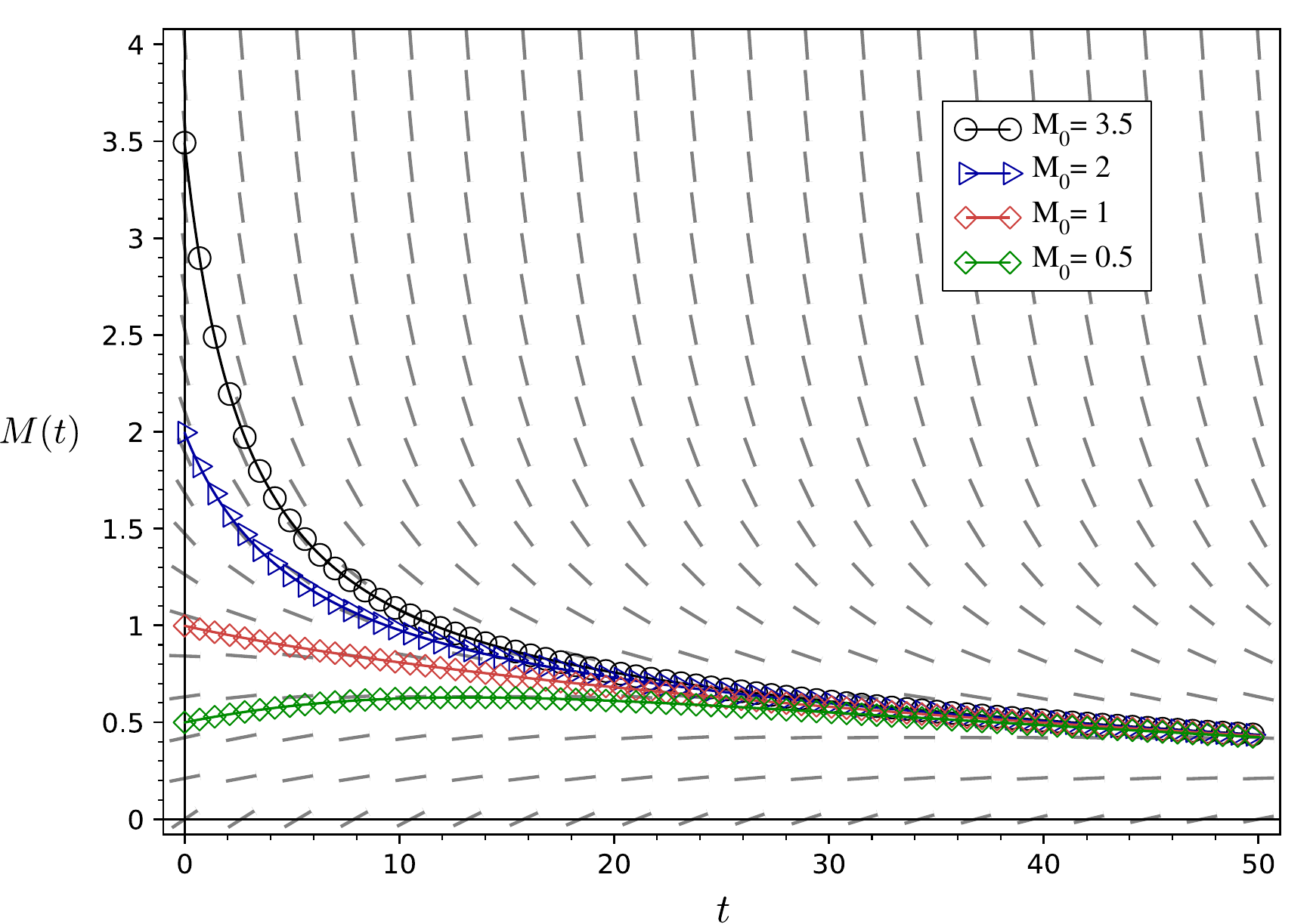}
    \caption{Dark energy era. It is considered $a = 0.1$, $b = 0.075$ and $\delta_\Lambda = 0.025$ and some values for initial mass per unit length $M_0$.}
    \label{fig:Mxt-field_dark}
\end{figure}

We analyze now the dark energy era by applying the equation \eqref{eq:Stefan-Boltzmann_evoluc} with $i = \Lambda$. The non-linear differential equation obtained is

\begin{equation} \label{eq:eq-dif-M_dark}
    \frac{dM_+}{dt} = - aM_+^2 + be^{-\delta_\Lambda t}M_+^{\frac{2}{3}}
\end{equation}
where

\begin{equation} \label{eq:delta_Lambda}
    \delta_\Lambda = 4\left(\frac{8\pi G}{3}\rho_{\Lambda_0}\right)^{\frac{1}{2}}.
\end{equation}
The numerical solution to the equation \eqref{eq:eq-dif-M_dark} is shown in figure \ref{fig:Mxt-field_dark}.

We remember that the choice of parameters is made with the objective of producing a clear plot. The constants of the problem are expressed in arbitrary units. The parameters introduced in equation \eqref{eq:eq-dif-M_dark} are also written in arbitrary units. The figure \ref{fig:Mxt-field_dark} can be understood as a confirmation of the predictions made in this paper. Since we make the assumption that the heat capacity of the black rope is a positive quantity, we expect that a state of thermal equilibrium shall be reached and therefore an equilibrium mass shall be reached. Considering the figure \ref{fig:Mxt-field_dark}, it is possible to observe that the black string mass per unit length decreases more slowly than in the previous cases which means that the evaporation is less intense. If we consider all the three cases investigated in this section we notice that in the case of matter era, the black string evaporation is more intense than the others and we observe a similar behavior when we compare the cases of radiation era, figure \ref{fig:Mxt-field_rad}, and dark energy era, figure \ref{fig:Mxt-field_dark}.

\section{Conclusions}

This paper represents a conjecture regarding the presence of the black string in the universe as we know it considering its eras. This conjecture is interesting to situate the black string model in the temporal evolution of the universe. In principle, the black string subjected to the universe as we currently understand it would be an object in thermal equilibrium with the universe, with respect to current measurements of cosmic microwave background radiation temperature. This interpretation refers to figure 2 in which it is clearly seen that as time passes, the mass per unit length assumes a constant value. This means that the rate of energy exchange between the black string and the reservoir (CMB) has ceased and thermal equilibrium has been reached. The study considering the change in reservoir temperature as time passes performed in sections \ref{subsec:era_rad}, \ref{subsec:era_mat} and \ref{subsec:era_dark} shows that the mass values decrease rapidly indicating a rapid evaporation of the black string as can be seen by investigating figures \ref{fig:Mxt-field_rad}, \ref{fig:Mxt-field_mat} and \ref{fig:Mxt-field_dark}. These conjectures are important for an understanding of the thermal behavior of objects representing singularities and for using a method for such studies.

\begin{acknowledgments}
The authors thank to Brazilian agency CAPES - Coordenação de Aperfeiçoamento de Pessoal de Nível Superior (Coordination for the Improvement of Higher Education Personnel) for financial support.
\end{acknowledgments}

\bibliography{Bibliography} 

\end{document}